\documentclass[showpacs,twocolumn,aps]{revtex4-1}

\usepackage{graphicx}
\usepackage{amssymb}
\usepackage{amsfonts}
\usepackage{amsmath}
\usepackage{amsbsy}
\usepackage{natbib}
\usepackage{comment}
\usepackage{color}
\usepackage{float}

\renewcommand{\vec}[1]{\mbox{\boldmath$\mathrm{#1}$}}
\newcommand{\be}{\begin{equation}}
\newcommand{\ee}{\end{equation}}
\newcommand{\ben}{\begin{eqnarray}}
\newcommand{\een}{\end{eqnarray}}

\begin{document}

\title{Conversion of electronic to magnonic spin current at heavy-metal magnetic-insulator interface}

\author{Xi-guang Wang$^{1,2}$}

\author{Zhi-xiong Li$^{1}$}

\author{Zhen-wei Zhou$^{1}$}

\author{Yao-zhuang Nie$^{1}$}

\author{Qing-lin Xia$^{1}$}

\author{Zhong-ming Zeng$^{3}$}

\author{L. Chotorlishvili$^{2}$}

\author{J. Berakdar$^{2}$}

\author{Guang-hua Guo$^1$}
\email{guogh@mail.csu.edu.cn}

\affiliation{
	$^{1}$School of Physics and Electronics, Central South University, Changsha 410083, China \\
	$^{2}$Institut f\"ur Physik, Martin-Luther Universit\"at Halle-Wittenberg, 06099 Halle (Saale), Germany\\
	$^{3}$Suzhou Institute of Nano-tech and Nano-bionics, Chinese Academy of Sciences, Suzhou, 215123, China
}

\begin{abstract}
 Electronic spin current is convertible to magnonic spin current via the creation or annihilation of thermal magnons at the interface of a magnetic insulator and a metal with a strong spin-orbital coupling. So far this phenomenon was evidenced in the linear regime. Based on analytical  and full-fledged numerical results for the  non-linear regime we demonstrate that the generated thermal magnons or magnonic spin current in the insulator is asymmetric with respect to the charge current direction in the metal and exhibits a nonlinear dependence on the charge current density, which is explained by the tuning effect of the spin Hall torque and the magnetization damping. The results are also discussed in light of  and are in line with recent experiments  pointing to a new way of non-linear manipulation of spin with electrical means.
\end{abstract}
\pacs{72.25.Mk, 75.30.Ds, 75.40.Gb, 75.70.Cn}

\date{\today}

\maketitle

Magnon, the quanta of the collective magnetic excitations in  a magnetically ordered materials carries a spin angular momentum $ \hbar $, directed opposite to the local magnetic moment. Thus, a flow of magnons during non-equilibrium magnonic excitations  implies an angular momentum flow, or a   spin current  \cite{GEWBauer2012, YKajiwara2010,KUchida2010, DHinzke2011}. This magnonic spin current can be employed to carry, transport, and process information \cite{GEWBauer2012, YKajiwara2010, VVKruglyak2010, BLenk2011, AVChumak2015}, as well as to generate a spin torque acting on the local magnetic moment that can be exploited to drive magnetization dynamics \cite{XJia2011, CBi2014} and  magnetic  domain walls  \cite{DHinzke2011, XgWang2012, PYan2011,WJiang2013, FSchlickeiser2014,seyyed}. Employing magnetic insulators reduces significantly  Joule losses due the  magnon current and hence may save energy. An adverse point however  is that
 direct magnonic excitations via a magnetic field implies heat dissipation associated with the magnetic field generation and a slow operation, as the production of short magnetic pulses (say pico or sub picosecond) is a challenge. Hence, ways to trigger magnonic spin current electrically are highly advantageous.
 One way to achieve that is via the inter-conversion between the electronic and magnonic spin currents at interfaces  \cite{YKajiwara2010, SSLZhang2012, JLi2016, HWu2016, LJCornelissen2015, CCiccarelli2015, STBGoennenwein2015}. Till now, basically two methods accomplish the charge to magnon  conversion. The first one was formulated by Kajiwara \textit{et al} \cite{YKajiwara2010}: Considering a heavy metal on a magnetic insulator, say a Pt stripe on YIG, due to the self-sustained magnetization oscillation induced by the spin-transfer torque, the electronic spin current generated through the spin Hall effect in Pt stripe, is converted to a magnonic spin current in YIG. This method requires a large charge current density in Pt to overcome the magnetization damping and the effect is highly nonlinear. An alternative proposal was suggested theoretically by S. Zhang \textit{et al} \cite{SSLZhang2012} and subsequently approved experimentally for devices with different geometries \cite{JLi2016, HWu2016, LJCornelissen2015, CCiccarelli2015, STBGoennenwein2015}: Instead of exciting spin waves through the spin transfer torque,  thermal magnons are created or annihilated based on the electron-magnon interaction directly at the normal metal-magnetic insulator interface. The non-equilibrium thermal magnons diffuse away or towards the interface, generating thus a magnonic spin current. This mechanism is linear in the charge current. The generated magnonic spin current can be controlled by tuning the magnetization direction.

In spite of the formal difference, both methods have certain similarity, as in both cases the underlying physical mechanism is based on the spin Hall torque (or spin-orbital torque). In the first method the spin torque plays a key role for the ferromagnetic resonance. In the second method the spin torque is used to create or annihilate thermal magnons, and the effects  depends strongly on the temperature. Here
 we study in detail the creation (annihilation) of the thermal magnons at the Pt / YIG interface and the magnon diffusion in YIG. Our theoretical micromagnetic approach is appropriate to the discussed effects and explains recent experimental observations \cite{JLi2016, HWu2016, LJCornelissen2015}. We found that the magnonic spin current is asymmetric with respect to the direction of the electric current. Besides, the converted magnonic spin current depends non-linearly on the electrical current density.

\begin{figure}[h]
	\centering
	\includegraphics[width=0.49\textwidth]{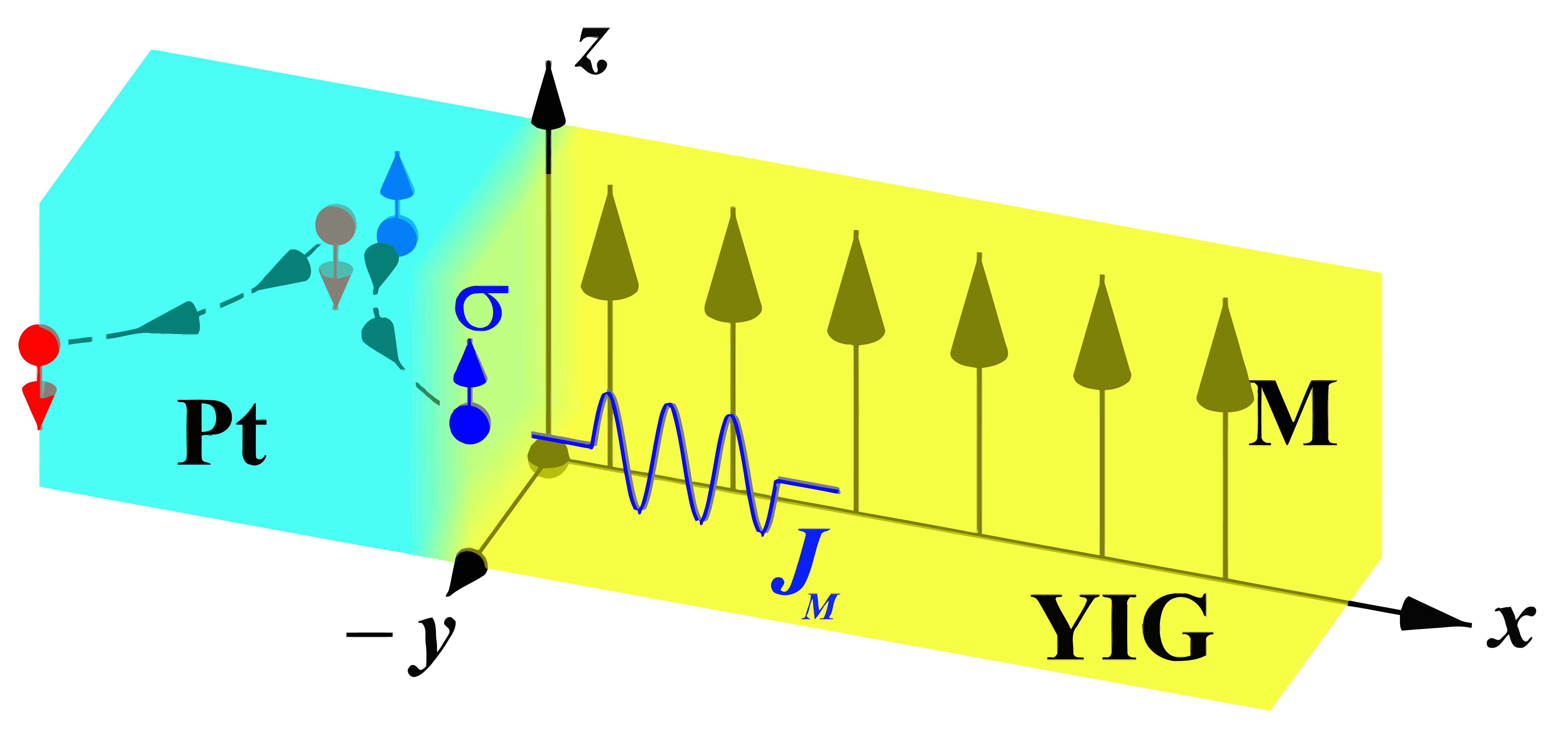}
	\caption{(color online). Schematics of the model. Magnetization of the YIG layer is oriented collinear with the +\textit{z} axis. The pure spin current with the spin polarization $ \vec{\sigma} $  is generated in the Pt layer through the spin Hall effect, and the electric current $ j_{\mathrm{Pt}} $. The electronic spin current is converted to a magnonic spin current $ J_M $ at the Pt / YIG interface via the spin-Hall torque.}
	\label{fig_1}
\end{figure}

We consider a one-dimensional Pt / YIG structure shown in Fig. \ref{fig_1}. The Pt layer is attached to the end of the YIG magnet. Due to spin-orbital coupling, biased Pt exhibits a spin Hall effect exerting
 a spin torque proportional to $ \vec{M} \times \vec{\sigma} \times \vec{M} $ on YIG \cite{LLiu2012, KGarello2013, AHoffmann2013}. Here $ \vec{\sigma} $ is the polarization of the spin current and $ \vec{M} $ is the magnetization of the YIG. The electronic spin current in Pt can be converted to a magnonic spin current in the YIG  either by driving a ferromagnetic resonance \cite{YKajiwara2010, RHLiu2013, AGiordano2014}, or by creating (or annihilating) thermal magnons \cite{SSLZhang2012, JLi2016, HWu2016}. In the first method, conversion is more efficient when $ \vec{\sigma} $ is perpendicular to the equilibrium magnetization, while in the second method conversion works best when $ \vec{\sigma} $ is parallel to the static magnetization (that means when $ \vec{\sigma} $  is perpendicular to the thermally activated dynamic component of the magnetization).

\begin{figure}[h]
	\centering
	\includegraphics[width=0.49\textwidth]{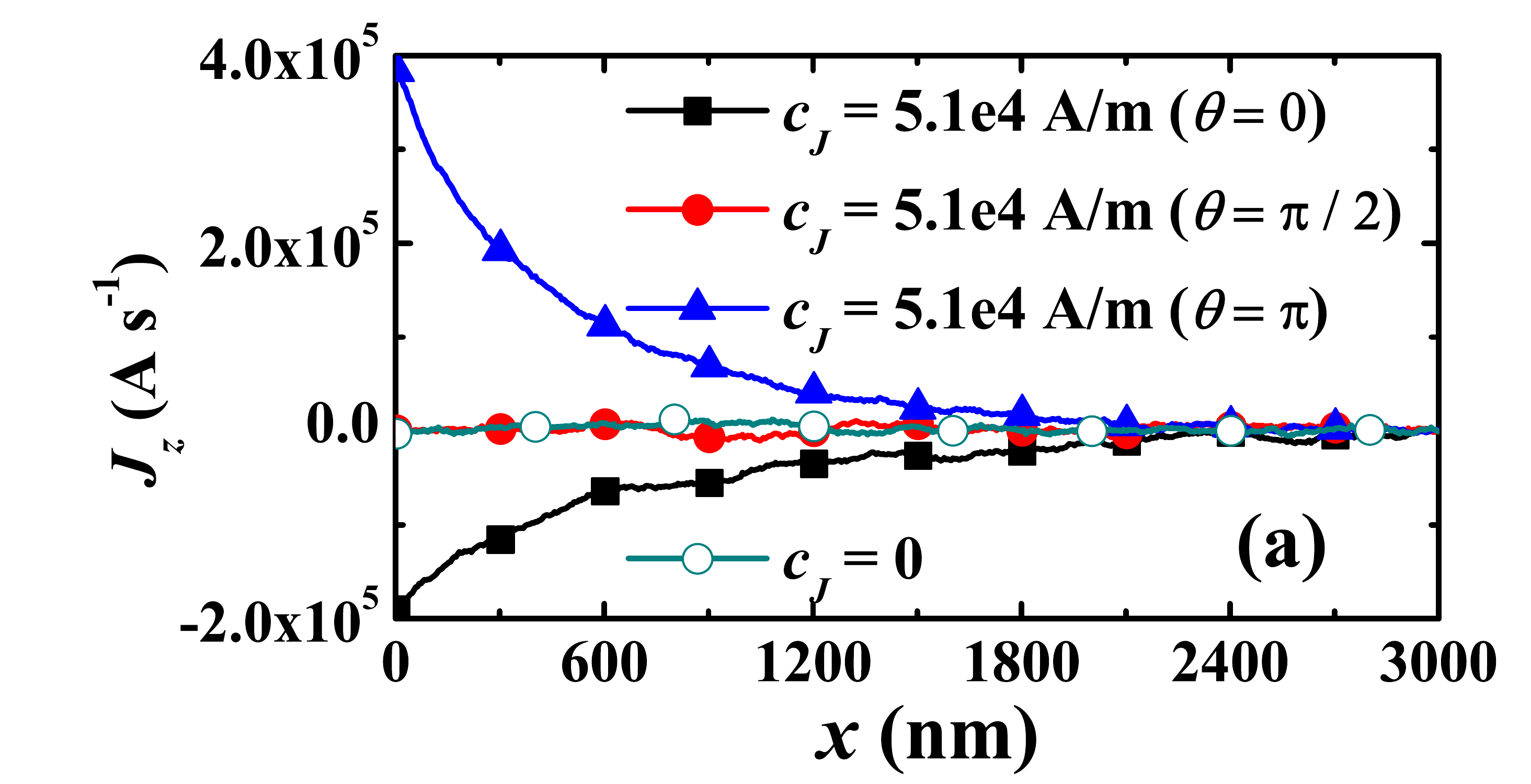}
	\includegraphics[width=0.49\textwidth]{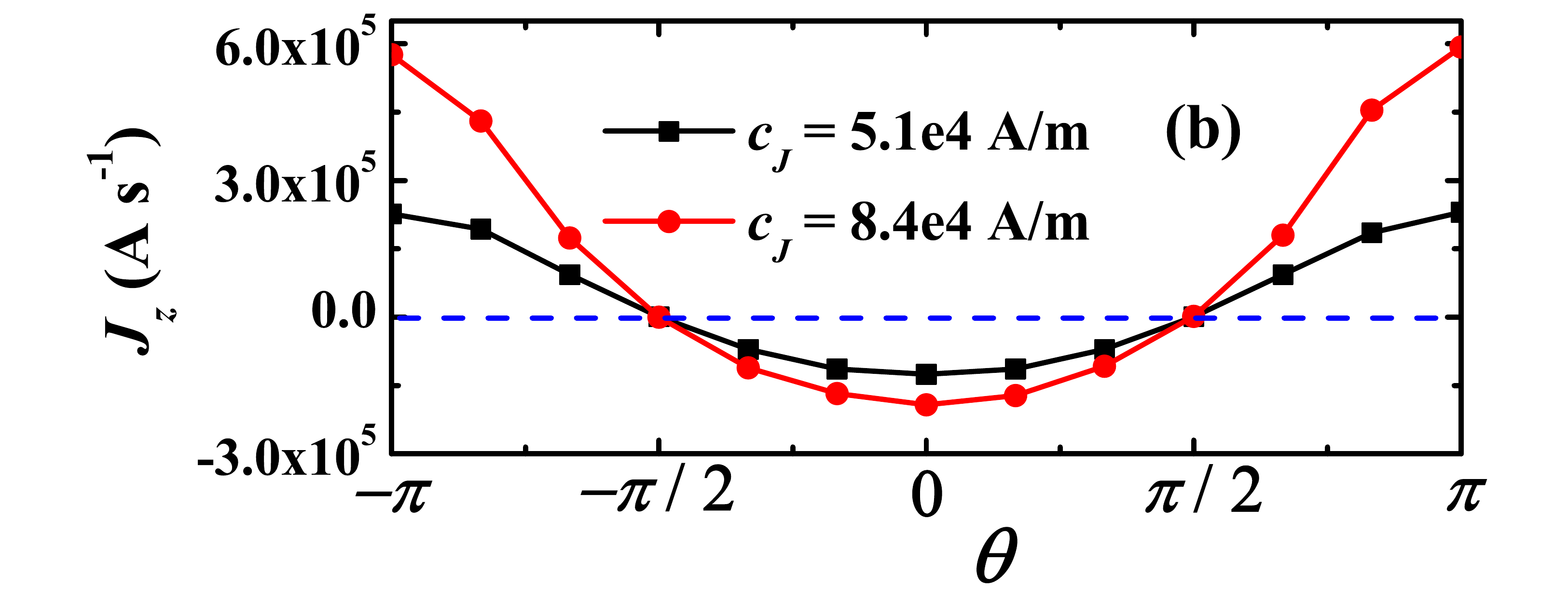}
    \includegraphics[width=0.49\textwidth]{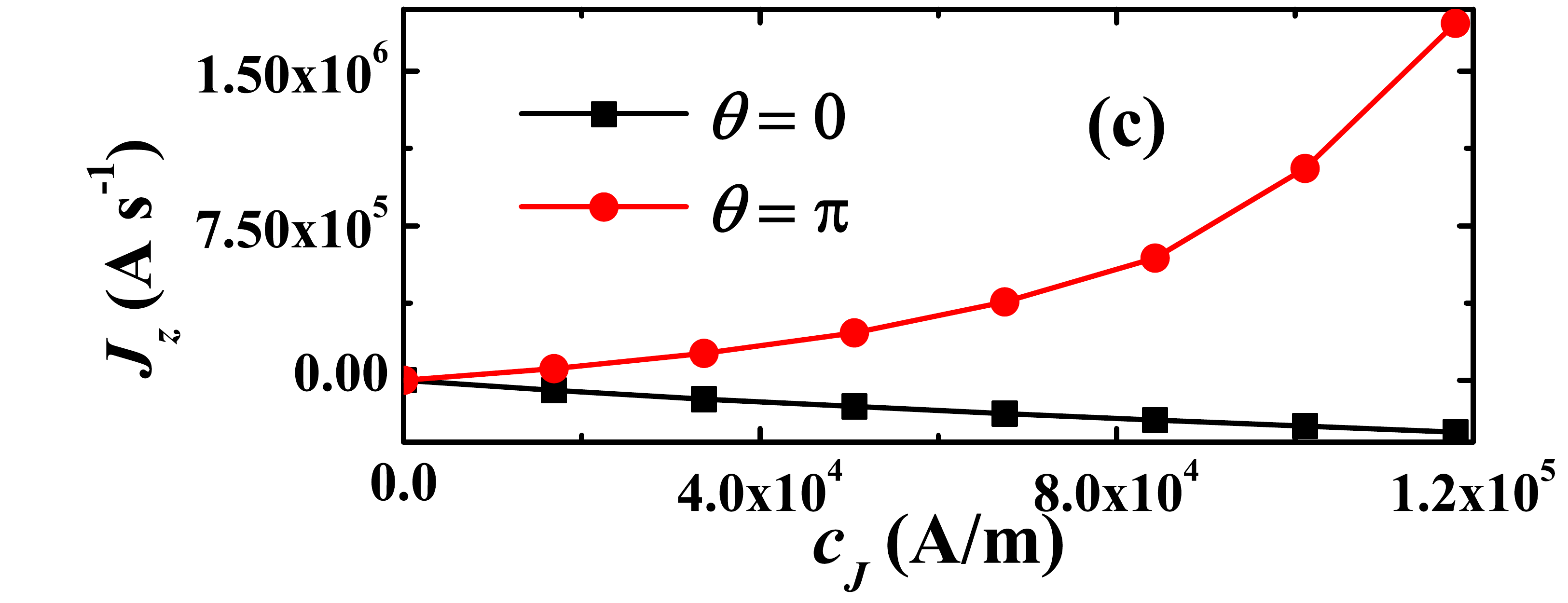}
	\caption{(color online). (a) Numerically simulated profile of the magnonic spin current $ J_z $ in YIG for the parameters $ c_J = 0 $ (open circles), $ c_J = 5.1 \times 10^4 $ A/m with $ \theta = 0 $ (squares), $ \theta = \pi /2 $ (solid circles) and $ \theta = \pi $ (triangles). The temperature is $ T = 25 $ K. (b) Averaged spin current $ J_z $ (in the region of $ x \le 500 $ nm) as a function of the angle $ \theta $ for $ c_{J} = 5.1 \times 10^4 $ A/m and $ 8.4 \times 10^4 $ A/m. (c) Averaged spin current $ J_z $ as a function of $ c_J $ for $ \theta = 0  $ and $ \pi $.}
	\label{fig_2}
\end{figure}

Here we address the case when the charge current density in Pt is very small and a self-sustained magnetization precession is not activated. The relevant mechanism in this case is the creation (annihilation) of thermal magnons at the Pt / YIG interface. Due to the non-equilibrium thermal magnon diffusion, the magnonic spin current flows along the \textit{x}-axis. The magnetization dynamics in YIG is governed by the stochastic Landau-Lifshitz-Gilbert (LLG) equation supplemented by the spin Hall torque $ \tau_{\mathrm{SHE}} $ term \cite{JLGarcíaPalacios1998}:
\begin{equation}
\displaystyle \partial_t \vec{M} = - \gamma \vec{M} \times (\vec{H}_{\mathrm{eff}}+\vec{h}_{l})+ (\alpha / M_{s})\vec{M} \times \partial_t \vec{M} + \tau_{\mathrm{SHE}}.
\label{eq_1}
\end{equation}
Here $ M_s $ is the saturation magnetization, $ \gamma $ is the gyromagnetic ratio, and $ \alpha $ is the phenomenological Gilbert damping constant. The effective field $ \vec{H}_{\mathrm{eff}} $ consists of the exchange field and the applied external magnetic field, $ \vec{H}_{\mathrm{eff}}=\frac{2 A}{\mu_0 M_s^2} \frac{\partial^2 \vec{M}}{\partial x^2} + H_z \vec{z} $. Here $ A $ is the exchange stiffness and $ H_z $ is the external magnetic field applied along the \textit{z}-direction. The thermal random magnetic field $   \langle h_{l,i}(x,t) h_{l,j}(x',t') \rangle = \frac{2 k_{B} T \alpha}{\gamma M_{s} V}\delta_{ij} \delta(x-x') \delta(t-t') $ is characterized by the correlation function of the white noise. Here $ k_B $ is the Boltzmann constant, $ V $ is the volume and $ T $ is the temperature. The Pt / YIG interfacial spin Hall torque $ \tau_{\mathrm{SHE}} $ created by the electronic current flowing in the Pt layer can be expressed as $ \tau_{\mathrm{SHE}}= \frac{\gamma c_J}{M_s} \vec{M} \times \vec{\sigma} \times \vec{M} $.
Here the coefficient $ c_J $ is proportional to the electronic current density $ J_{\mathrm{Pt}} $ in the Pt layer. The polarization of the spin current $ \vec{\sigma} $ satisfies $ \vec{\sigma} = \vec{x} \times \vec{j_{\mathrm{Pt}}} $, and $ \vec{j_{\mathrm{Pt}}} $ is the unit vector of the electronic current \cite{LLiu2012, KGarello2013, AHoffmann2013}.

In the modeling of the creation / annihilation of thermal magnons at the Pt/YIG interface and magnon diffusion in YIG, the length of YIG is chosen to be 3000 nm and the cell size is 5 nm. The material parameters of YIG used in simulation are: $ M_s = 1.4 \times 10^5 $ A/m, $ A = 3 \times 10^{-12} $ J/m, and $ \alpha = 0.005 $ \cite{YSun2013}. The YIG is initially magnetized to saturation and the magnetization is aligned along the +\textit{z} axis parallel to the large external magnetic field $ H_z = 4 \times 10^5 $ A/m. The spin Hall torque act in the vicinity of the interface, and its coefficient $ c_J $ is of the order of  (0, $ 1.2 \times 10^5 $ A/m): This value of $ c_J $ is small enough to exclude the reorientation of magnetization and the self-sustained auto-oscillation induced by the spin Hall torque. Influence of the relative angle $ \theta $ between the electron spin and the static magnetization can be investigated by controlling the direction of the current density vector $ \vec{j_{\mathrm{Pt}}} $ and the polarization of the spin current $ \vec{\sigma} = (0, \mathrm{sin}\theta, \mathrm{cos}\theta) $. The temperature $ T $ is uniform in the YIG and far below the Curie temperature.

In order to quantify the magnonic spin current $ J_s $, we utilize the standard definition $ J^{\alpha}_s = l_A \varepsilon_{\alpha \mu \nu} \langle M_{\mu} \partial_{x} M_{\nu} \rangle $ \cite{YKajiwara2010}. Here $ \alpha $ defines polarization of the magnonic spin current, $ l_A = 2 \gamma A / (\mu_0 M^2_s)$ and $ \varepsilon_{\alpha \mu \nu} $ is the Levi-Civita antisymmetric tensor. As the static magnetization is directed along the +\textit{z} axis parallel to the external magnetic field $ H_z $, the thermally activated transversal magnetization components $ M_x $, $ M_y $ contribute only to the magnonic spin current $ J_s^z $. Two other components of the magnonic spin current $ J_s^x $ and $ J_s^y $ are zero. Taking into account that the magnon polarization is opposite to the local magnetization, we set $ J_z = -J^z_s $ in order to have a positive spin current $ J_z $ for magnons propagating along the +\textit{x} direction.

\begin{figure}[h]
	\centering
	\includegraphics[width=0.49\textwidth]{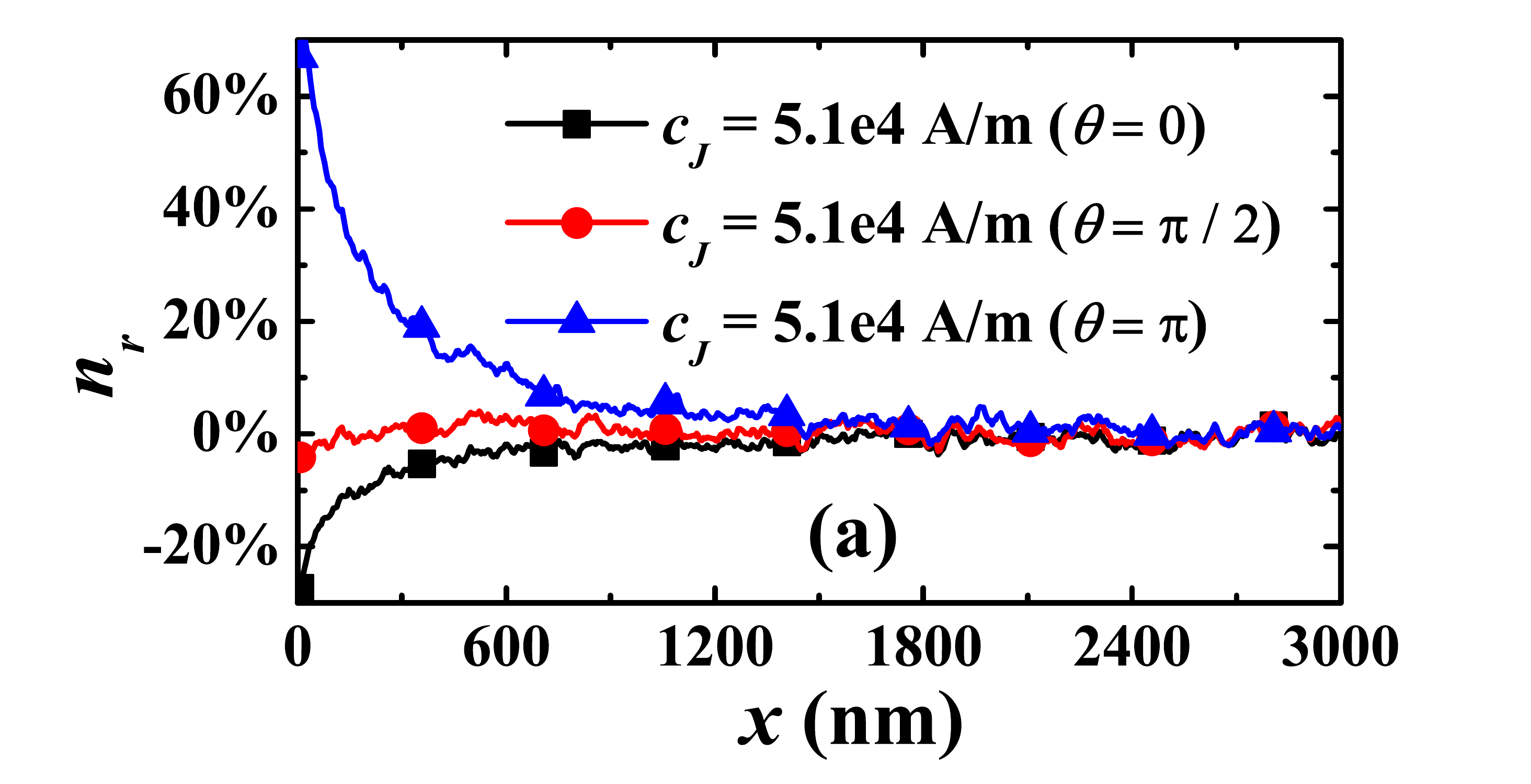}
	\includegraphics[width=0.49\textwidth]{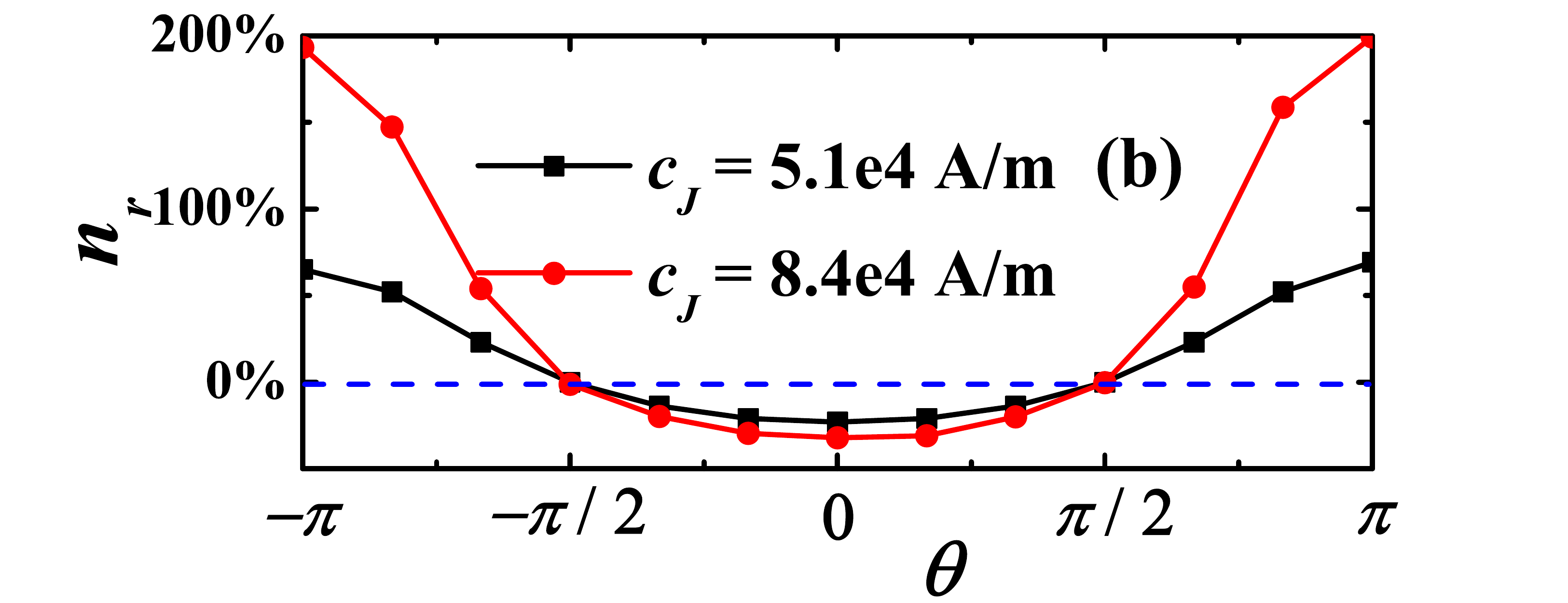}
	\includegraphics[width=0.49\textwidth]{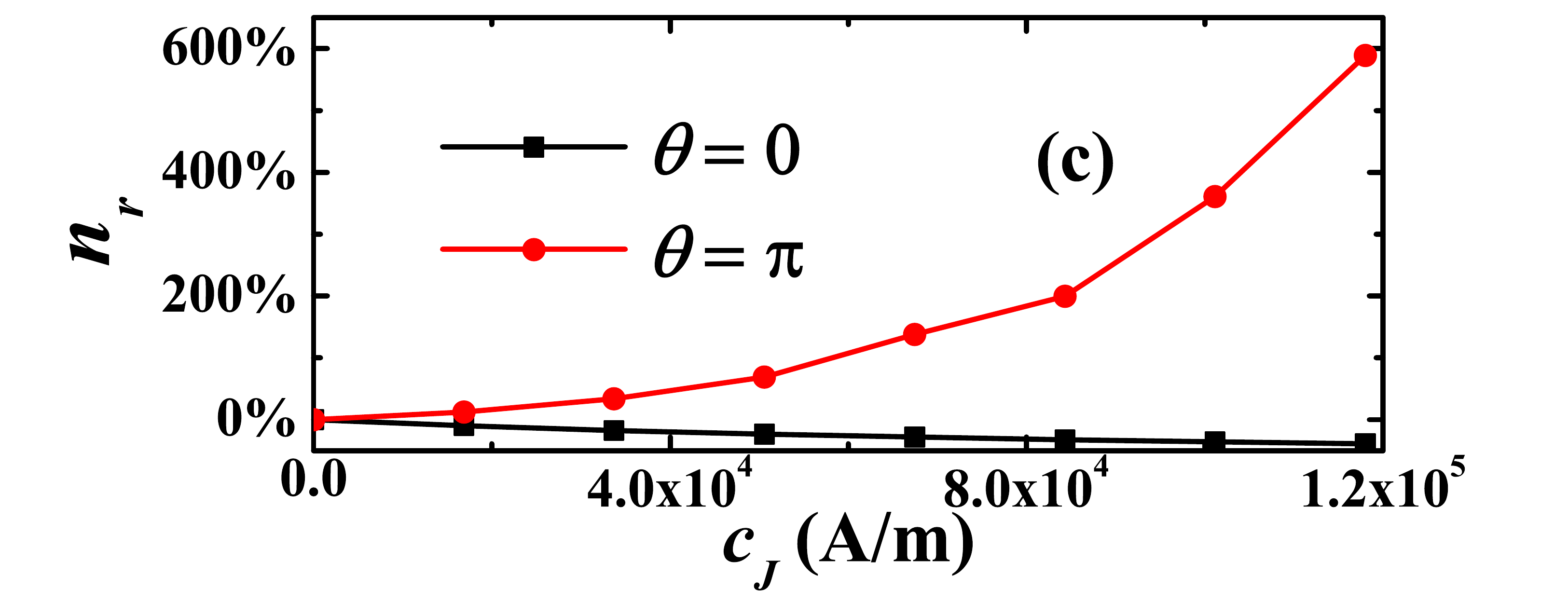}
	\caption{(color online). (a) Numerically simulated profile of relative magnon density $ n_r $ for the parameters $ c_J = 5.1 \times 10^4 $ A/m with $ \theta = 0 $ (squares), $ \theta = \pi /2 $(circles) and $ \theta = \pi $ (triangles). The temperature is $ T = 25 $ K. (b) Averaged $ n_r $ as a function of the angle $ \theta $ for $ c_J = 5.1 \times 10^4 $ A/m and $ 8.4 \times 10^4 $ A/m. (c) Averaged $ n_r $ as a function of $ c_J $ for $ \theta = 0  $ and $ \pi $.}
	\label{fig_3}
\end{figure}

Figure \ref{fig_2}(a) shows the spatial distribution of the magnonic spin current density $ J_z $ in YIG with and without the spin Hall torque being applied at the Pt / YIG interface (\textit{x} = 0). Here, \textit{T} is taken to be 25 K. In the absence of the spin Hall torque ($ c_J = 0 $), the uniform temperature can not generate magnon current and $ J_z = 0 $. When the spin Hall torque $ c_J = 5.1 \times 10^4 $ A/m is applied, the electronic current is converted into a magnonic spin current. The effect is sensitive  to the angle $ \theta $. When the electron spin polarization is parallel to the static magnetization ($ \theta $ = 0), the magnonic spin current $ J_z $ is negative, meaning that the nonequilibrium thermal magnons diffuse toward the interface. The spin current is exponentially damped along the \textit{x} direction. The magnonic spin current $ J_z $ turns positive when $ \theta = \pi $, and its value is much larger than in the case $ \theta = 0 $. If the electron polarization is perpendicular to the static magnetization ($ \theta = \pi / 2 $), the magnonic spin current is zero. The slight variations in these curves are caused by  thermal effects. The detailed connection between the magnonic spin current density $ J_z $ and the electron spin polarization angle $ \theta $ is shown in Fig. \ref{fig_2}(b). The dependence of the magnonic current $ J_z $ on the angle $ \theta $ mimics the profile of a harmonic function. Conversion of the electronic spin current into the magnonic spin current is maximal for $ \theta = 0 $ and $ \pm \pi $. However, considering  the polarization of the electronic spin and the direction of the electric current in Pt one notices an asymmetry.
 A larger magnonic spin current is generated when the electron polarization is antiparallel to the static magnetization ($ \theta = \pm \pi $). This phenomenon becomes more prominent for higher electronic current density ($ c_J = 8.4 \times 10^4 $ A/m). The dependence of the magnonic spin current density $ J_z $ on the spin torque coefficient $ c_J $ is shown in Fig. \ref{fig_2}(c) for $ \theta = 0  $ and $ \pi $. The magnonic spin current increases with the electronic current in Pt. But $ J_z $ is not linearly proportional to $ c_J $ as shown in Ref. \cite{JLi2016, LJCornelissen2015}. The nonlinear effect enhances at high electronic current density $ c_J $ and particularly  for $ \theta = \pi $, while for $ \theta = 0 $ the effect is less pronounced.

The conversion between the electronic and the magnonic spin currents at the normal metal-magnetic insulator interface was studied in Ref. \cite{SSLZhang2012, JLi2016, HWu2016, LJCornelissen2015, LJCornelissen2016}. For $ \theta = 0 $, the electron spins are antiparallel to the thermal magnon's spins oriented opposite to the local magnetization, and magnons are annihilated due to the transfer of angular momentum. While for $ \theta = \pi $, magnons are created. The creation or annihilation of the thermal magnons at the Pt / YIG interface lead to positive or negative magnon accumulation. The magnons start to diffuse  then away or towards the interface. When the electron spin is perpendicular to the spin of the thermal magnon, $ \theta = \pi / 2 $, the spin transfer between the electron and the  thermal magnons is totally suppressed. Therefore, the  generated magnonic spin current should exhibit a symmetry with respect to the direction of the electronic current and should depend linearly on the electronic current density \cite{SSLZhang2012}.

 For a deeper  understanding of the conversion process we further calculate the thermal magnon density $ n $ and its distribution. The thermal magnon number $ n $ is quantified by the squared dimensionless transversal magnetization components averaged over time $ n = \rho V M_s/(2 g \mu_B ) $, where $ \rho = \langle m^2_x + m^2_y \rangle $ with $ m_x = M_x / M_s $ and $ m_y = M_y / M_s $ \cite{AGGurevich1996} .  $ \mu_B $ is the Bohr magneton. In Fig. \ref{fig_3}(a), the spatial distributions of the relative nonequilibrium magnon density $ n_r = (n - n_0) / n_0 $ are plotted. Here $ n_0 $ is the equilibrium magnon density when $ c_J = 0 $. We clearly see that the thermal magnons are created or annihilated at the Pt / YIG interface ($ x = 0 $) and the nonequilibrium magnons diffuse through the YIG when the  electronic current $ c_J $ is applied. The number of created magnons  is  larger than the number of annihilated magnons. This imbalance leads to the asymmetry of the magnonic spin current with respect to the direction of the charge current. Fig. \ref{fig_3}(a)   indicates clearly that there is no creation (annihilation) of thermal magnons when the electron spin polarization is perpendicular to the static magnetization, i.e. $ \theta = \pi / 2 $. The nonequilibrium thermal magnon density as a function of the angle $ \theta $ and spin torque coefficient $ c_J $ is shown in Fig. \ref{fig_3}(b) and (c), respectively. We clearly see the characteristic asymmetry of the nonequilibrium thermal magnons with respect to the direction of the electrical current and the nonlinear dependence on the current density.

Our micromagnetic simulations results are in good agreement with the experimentally observed data. In the experiment, the nonequilibrium thermal magnons created or annihilated at the Pt / YIG interface were detected at another Pt strip through the inverse spin Hall effect \cite{JLi2016, HWu2016, LJCornelissen2015, LJCornelissen2016}. The detected electrical signal depends on the angle between the electronic spin polarization and the static magnetization in the form $ \sim \mathrm{cos}^2 \theta $. This experimental result is consistent with our simulations. In our case the dependence of the magnonic current $ J_z $ on the angle $ \theta $ amounts to the profile of a harmonic function.

The asymmetry of the magnonic spin current with respect to the direction of the electric current was also observed in the experiment. An explanation of this asymmetry was formulated in terms of the spin Seebeck effect: The joule heating in the Pt layer generates a temperature gradient at the Pt / YIG interface and a positive magnon accumulation leads to the flow of magnons away from the interface \cite{JLi2016,LJCornelissen2015}.


 In our theoretical model the applied thermal bias is uniform and the spin Seebeck effect is not relevant. Therefore, the observed asymmetry and the nonlinear effects exposed here point to a different mechanism.

In our model the spin-Hall torque plays a dual role in the conversion between the electronic and the magnonic spin currents. On the one hand, the spin-Hall torque directly creates or annihilates thermal magnons at the interface by acting on the dynamic components of the magnetization. This conversion mechanism leads to the linear dependence of magnonic current on the electric current, as indicated in Ref. \cite{SSLZhang2012}. The spin-Hall torque   modifies linearly the magnetization damping. It increases or decreases the effective damping. Depending on the direction of the electric current, the spin-Hall torque enhances or attenuates the thermally activated oscillations of the magnetization \cite{Ando}. On the other hand, as will be shown below, the converted magnonic current has a contribution proportional to the square of the electric current.

\begin{figure}[h]
	\centering
	\includegraphics[width=0.49\textwidth]{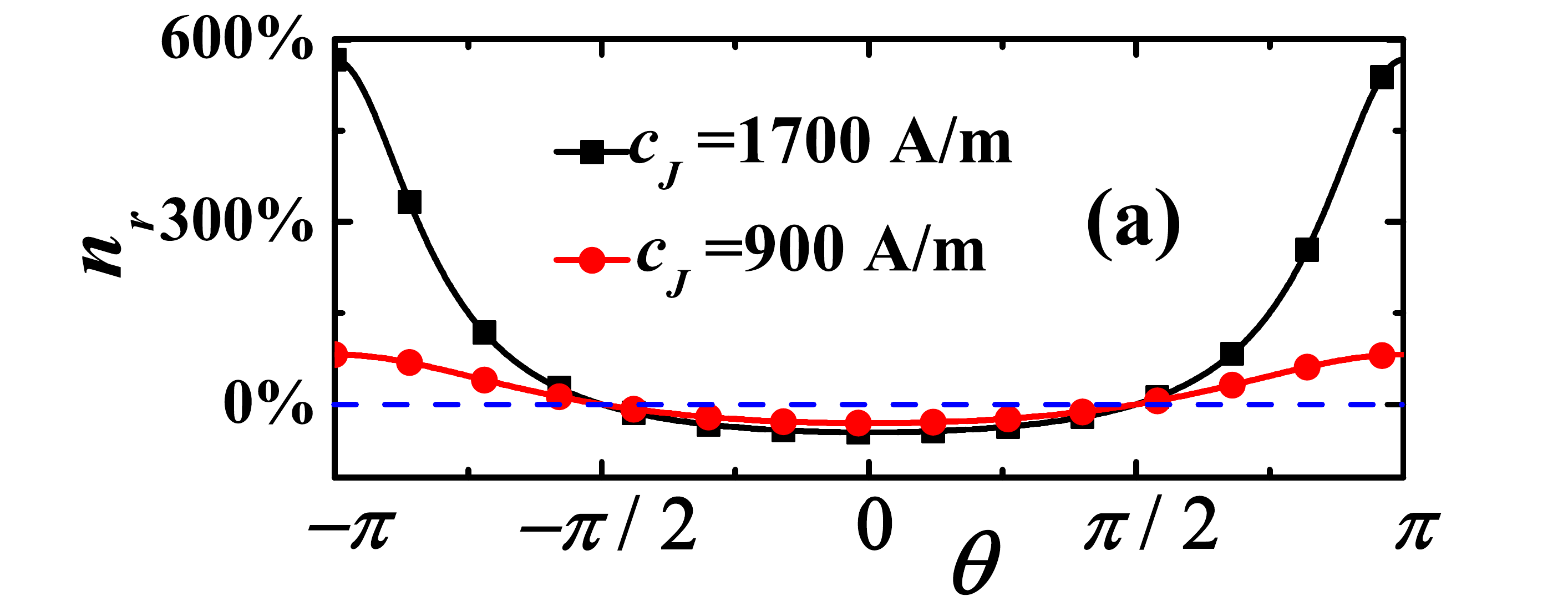}
	\includegraphics[width=0.49\textwidth]{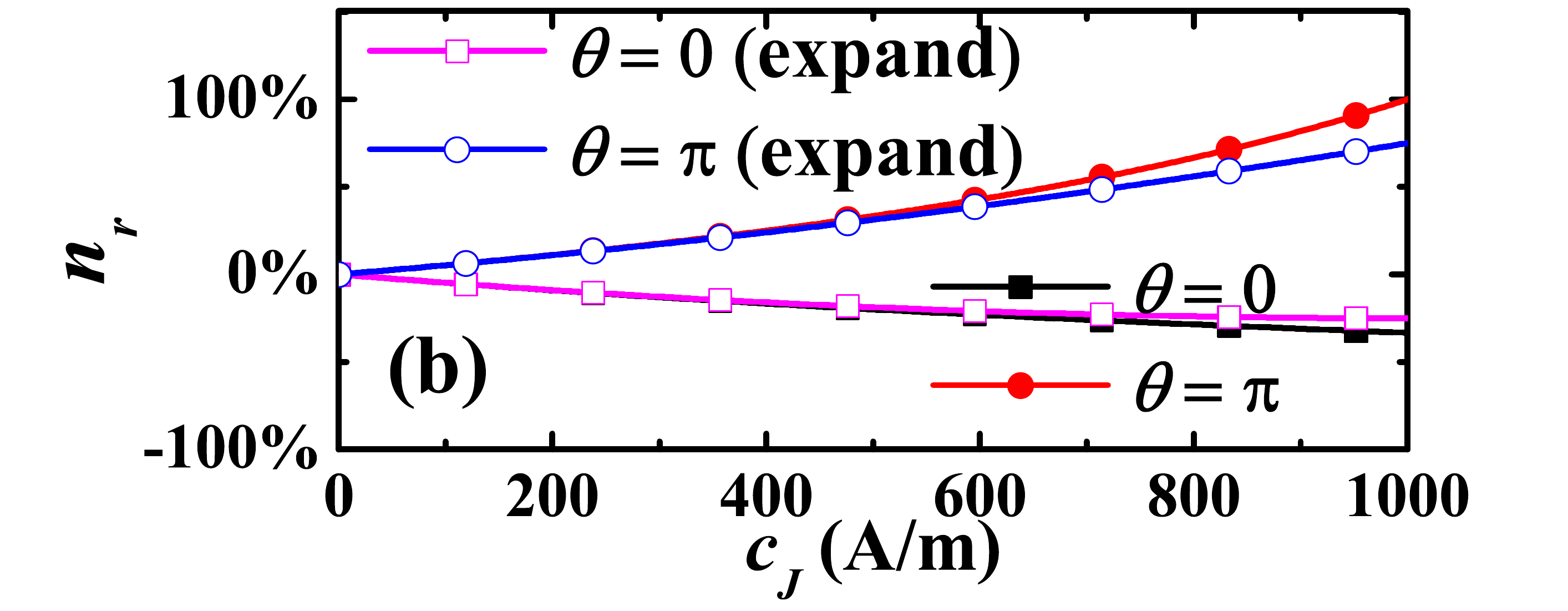}
	\caption{(color online). (a) Analytically calculated relative magnon density $ n_r $ as a function of the angle $ \theta $ when $ c_J = 1700 $ A/m and 900 A/m. (b)  Relative magnon density $ n_r $ as a function of $ c_J $ for $ \theta = 0 $ and $ \pi $ calculated using Eq. (\ref{eq_3}) (solid dots) and Eq. (\ref{eq_5}) (open dots).}
	\label{fig_4}
\end{figure}

 To describe the conversion of the electronic spin current into the magnonic spin current on the Pt / YIG interface, we construct a simple analytical model:  The thermally activated magnetization dynamics of a single macrospin can be expressed as,  $ \vec{M} = \vec{M}_0 + M_s [m_{x}(t) \vec{e}_{x}+m_{y}(t) \vec{e}_{y}] $. The static magnetization $ \vec{M}_0 = M_s \vec{e}_z $ is oriented parallel to the external magnetic field $ H_z $. We assume that the reorientation of the static magnetization $ \vec{M}_0 $ caused by the spin Hall torque is small enough and can therefore  be neglected. After substituting $ \vec{M} $ into the stochastic LLG equation Eq. (\ref{eq_1}) and introducing $ m_{\pm} (t) = m_{x} (t) \pm i m_{y} (t) $, we deduce a set of linearized equations
\begin{equation}
\begin{small}
\left\{
\begin{aligned}
(1-i \alpha)\frac{\partial m_{+}}{\partial t} &= (i \gamma H_z - \gamma c_J \mathrm{cos}\theta) m_{+} - i \gamma h_+   \\
(1+i \alpha)\frac{\partial m_{-}}{\partial t} &= (-i \gamma H_z - \gamma c_J \mathrm{cos}\theta) m_{-} + i \gamma h_{-}.   \\
\end{aligned}
\right.
\label{eq_2}
\end{small}
\end{equation}
The complex Langevin field $ h_{\pm} = \vec{h}_l (\vec{e}_x \pm i \vec{e}_y) $ has the correlator $ \langle h_+(t) h_{-}(t) \rangle = 2 \langle h_{l,i}(t) h_{l,i}(t') \rangle  $ \cite{SHoffman2013}. Eq. (\ref{eq_2}) is linear and can be integrated straightforwardly. We utilize the method described in \cite{Saitoh1}
presenting  for brevity the final result for the correlation function
  \begin{equation}
   \begin{small}
\langle m_i(t)m_j(0) \rangle = \sigma^2  \int e^{-i\omega t} \sum_{n}\chi_{in}(\omega, c_{J}) \chi_{jn}(-\omega, c_{J})\frac{d \omega}{2 \pi}.
 \label{eq_3}
 \end{small}
 \end{equation}
Here, $ i, j, n = x,y $, and $\chi(\omega)$ is the transverse dynamic magnetic susceptibility matrix:
 \begin{equation}
  \begin{small}
\displaystyle \chi(\omega, c_{J}) = \frac{1}{(\omega_{H}-i\alpha \omega)^2 + (\omega_{c} - i \omega)^2} \left( \begin{matrix} \omega_H-i\alpha \omega & \omega_{c}-i\omega \\ i\omega-\omega_{c} & \omega_H-i\alpha \omega \end{matrix} \right).
 \label{eq_4}
  \end{small}
 \end{equation}
 The following notations are introduced: $ \omega_{c} = \gamma c_J \mathrm{cos}\theta$, $ \omega_{H} = \gamma H_z $, and  $ \sigma^{2} = 2\alpha \gamma k_B T / M_sV $.
In the high temperature limit Eq. (\ref{eq_3}) further simplifies and we infer
 \begin{equation}
 \setlength{\abovedisplayskip}{3pt}
 \setlength{\belowdisplayskip}{3pt}
    \begin{split}
 &\langle m_{x,y}(0)m_{x,y}(0) \rangle = \\
 &\frac{\gamma k_{B}T}{M_{s}V\omega_{H}}\bigg(1-\frac{\gamma c_{J}}{\alpha\omega_{H}}\cos(\theta)+\big(\frac{\gamma c_{J}}{\alpha\omega_{H}}\big)^{2}\cos(\theta)^{2}\bigg).
  \label{eq_5}
  \end{split}
 \end{equation}
Here $\frac{\gamma c_{J}}{\alpha\omega_{H}}$ is the small parameter in the  series expansion. With the definition of the magnon density $ \rho = \langle m^2_x + m^2_y \rangle $
one can readily calculate the magnon number $ n = \rho V M_s/(2 g \mu_B ) $.
As we see from Eq. (\ref{eq_5}), due to the linear term, the spin Hall torque enhances or attenuates the magnon density depending on the direction of the electric current. The effect of the spin Hall torque on the effective damping is linear $ \alpha_{\mathrm{eff}} = \alpha + c_J \mathrm{cos}\theta / H_z $, while the thermal magnon density depends on the quadratic term $\big(\frac{\gamma c_{J}}{\alpha\omega_{H}}\big)^{2}\cos(\theta)^{2}$. Thus, the spin Hall torque leads to an asymmetry and nonlinear effects in the creation or annihilation of thermal magnons at the heavy-metal magnetic-insulator interface. The analytical result from Eqs. (\ref{eq_3}) and (\ref{eq_5}) reproduces the nonlinear effects observed in micromagnetic simulations, as shown in Fig. \ref{fig_4}.

In summary, our theoretical results  explain  recent experiments on the conversion of  charge current to  magnonic spin current at the Pt / YIG interface. In particular, we discovered a relation between the  current  directional asymmetry and the nonlinear dependence of the magnonic spin current on the charge current.
Our results and the interpretations are essential elements towards the generation of non-linear, large magnonic current density via electrical means.

This work was supported by the National Natural Science Foundation of China under Grants No. 11674400 and No. 11374373 as well as by the German Science Foundation, DFG under SFB 762.




\begin{thebibliography}{29}

\bibitem{GEWBauer2012}
G. E. W. Bauer, E. Saitoh, and B. J. van Wees, Nat. Mater. {\bf 11}, 391 (2012).

\bibitem{YKajiwara2010}
Y. Kajiwara, K. Harii, S. Takahashi, J. Ohe, K. Uchida, M. Mizuguchi, H. Umezawa, H. Kawai, K. Ando, K. Takanashi, S. Maekawa, and E. Saitoh, Nature (London) {\bf 464}, 262 (2010).

\bibitem{KUchida2010}
K. Uchida, J. Xiao, H. Adachi, J. Ohe, S. Takahashi, J. Ieda, T. Ota, Y. Kajiwara, H. Umezawa, H. Kawai, G. E. W. Bauer, S. Maekawa, and E. Saitoh, Nat. Mater. {\bf 9}, 894 (2010).

\bibitem{DHinzke2011}
D. Hinzke and U. Nowak, Phys. Rev. Lett. {\bf 107}, 027205 (2011).

\bibitem{VVKruglyak2010}
V. V. Kruglyak, S. O. Demokritov, and D. Grundler, J. Phys. D: Appl. Phys. {\bf 43}, 264001 (2010).

\bibitem{BLenk2011}
B. Lenk, H. Ulrichs, F. Garbs, and M. M\"unzenberg, Phys. Rep. {\bf 507}, 107 (2011).

\bibitem{AVChumak2015}
A. V. Chumak, V. I. Vasyuchka, A. A. Serga, and B. Hillebrands, Nat. Phys. {\bf 11}, 453 (2015).

\bibitem{XJia2011}
X. Jia, K. Xia, and G. E. W. Bauer, Phys. Rev. Lett. {\bf 107}, 176603 (2011).

\bibitem{CBi2014}
C. Bi, L. Huang, S. Long, Q. Liu, Z. Yao, L. Li, Z. Huo, L. Pan, and M. Liu, Appl. Phys. Lett. {\bf 105}, 022407 (2014).

\bibitem{XgWang2012}
X.-g. Wang, G.-h. Guo, Y.-z. Nie, G.-f. Zhang, and Z.-x. Li, Phys. Rev. B {\bf 86}, 054445 (2012).

\bibitem{PYan2011}
P. Yan, X. S. Wang, and X. R. Wang, Phys. Rev. Lett. {\bf 107}, 177207 (2011).

\bibitem{WJiang2013}
W. Jiang, P. Upadhyaya, Y. Fan, J. Zhao, M. Wang, L.-T. Chang, M. Lang, K. L. Wong, M. Lewis, Y.-T. Lin, J. Tang, S. Cherepov, X. Zhou, Y. Tserkovnyak, R. N. Schwartz, and K. L. Wang, Phys. Rev. Lett. {\bf 110}, 177202 (2013).

\bibitem{FSchlickeiser2014}
F. Schlickeiser, U. Ritzmann, D. Hinzke, and U. Nowak, Phys. Rev. Lett. {\bf 113}, 097201 (2014).

\bibitem{seyyed} S.R. Etesami, L. Chotorlishvili, A. Sukhov, and J. Berakdar,
Phys. Rev. B \textbf{90},  014410 (2014); Appl. Phys. Lett. \textbf{107},  132402 (2015).

\bibitem{SSLZhang2012}
S. S. L. Zhang and S. Zhang, Phys. Rev. Lett. {\bf 109}, 096603 (2012).

\bibitem{JLi2016}
J. Li, Y. Xu, M. Aldosary, C. Tang, Z. Lin, S. Zhang, R. Lake, and J. Shi, Nat. Commun. {\bf 7} (2016).

\bibitem{HWu2016}
H. Wu, C. H. Wan, X. Zhang, Z. H. Yuan, Q. T. Zhang, J. Y. Qin, H. X. Wei, X. F. Han, and S. Zhang, Phys. Rev. B {\bf 93}, 060403 (2016).

\bibitem{LJCornelissen2015}
L. J. Cornelissen, J. Liu, R. A. Duine, J. B. Youssef, and B. J. van Wees, Nat. Phys. {\bf 11}, 1022 (2015).

\bibitem{CCiccarelli2015}
C. Ciccarelli, M. D. HalsKjetil, A. Irvine, V. Novak, Y. Tserkovnyak, H. Kurebayashi, A. Brataas, and A. Ferguson, Nat. Nano. {\bf 10}, 50 (2015).

\bibitem{STBGoennenwein2015}
S. T. B. Goennenwein, R. Schlitz, M. Pernpeintner, K. Ganzhorn, M. Althammer, R. Gross, and H. Huebl, Appl. Phys. Lett. {\bf 107}, 172405 (2015).

\bibitem{LLiu2012}
L. Liu, O. J. Lee, T. J. Gudmundsen, D. C. Ralph, and R. A. Buhrman, Phys. Rev. Lett. {\bf 109}, 096602 (2012).

\bibitem{KGarello2013}
K. Garello, I. M. Miron, C. O. Avci, F. Freimuth, Y. Mokrousov, S. Blugel, S. Auffret, O. Boulle, G. Gaudin, and P. Gambardella, Nat. Nano. {\bf 8}, 587 (2013).

\bibitem{AHoffmann2013}
A. Hoffmann, IEEE Trans. Magn. {\bf 49}, 5172 (2013).

\bibitem{RHLiu2013}
R. H. Liu, W. L. Lim, and S. Urazhdin, Phys. Rev. Lett. {\bf 110}, 147601 (2013).

\bibitem{AGiordano2014}
A. Giordano, M. Carpentieri, A. Laudani, G. Gubbiotti, B. Azzerboni, and G. Finocchio, Appl. Phys. Lett. {\bf 105}, 042412 (2014).

\bibitem{JLGarcíaPalacios1998}
J. L. García-Palacios and F. J. Lázaro, Phys. Rev. B {\bf 58}, 14937 (1998).

\bibitem{YSun2013}
Y. Sun, H. Chang, M. Kabatek, Y.-Y. Song, Z. Wang, M. Jantz, W. Schneider, M. Wu, E. Montoya, B. Kardasz, B. Heinrich, S. G. E. te Velthuis, H. Schultheiss, and A. Hoffmann, Phys. Rev. Lett. {\bf 111}, 106601 (2013).

\bibitem{LJCornelissen2016}
L. J. Cornelissen and B. J. van Wees, Phys. Rev. B {\bf 93}, 020403 (2016).

\bibitem{AGGurevich1996}
A. G. Gurevich and G. A. Melkov, Magnetization Oscillations and Waves (CRC, New York, 1996).

\bibitem{Ando}
K. Ando, S. Takahashi, K. Harii, K. Sasage, J. Ieda, S. Maekawa, and E. Saitoh Phys. Rev. Lett \textbf{101}, 036601 (2008)

\bibitem{SHoffman2013}
S. Hoffman, K. Sato, and Y. Tserkovnyak, Phys. Rev. B {\bf 88}, 064408 (2013).

\bibitem{Saitoh1}
J. Xiao, G. E. W. Bauer,. K.Uchida, E. Saitoh,and S. Maekawa  Phys. Rev. B \textbf{81}, 214418 (2010)


\end{thebibliography}
\end{document}